# Enhanced feedback performance in off-resonance AFM modes through pulse train sampling

Mustafa Kangül[1], Navid Asmari[1], Santiago H. Andany[1], Marcos Penedo[1], and Georg E. Fantner[1]

[1]Laboratory for Bio- and Nano-Instrumentation, Swiss Federal Institute of Technology Lausanne (EPFL), Lausanne CH-1015, Switzerland

## Abstract

Dynamic atomic force microscopy (AFM) modes that operate at frequencies far away from the resonance frequency of the cantilever (off-resonance tapping (ORT) modes) can provide high-resolution imaging of a wide range of sample types, including biological samples, soft polymers, and hard materials. These modes offer precise and stable control of vertical force, as well as reduced lateral force. Simultaneously, they enable mechanical property mapping of the sample. However, ORT modes have an intrinsic drawback: a low scan speed due to the limited ORT rate, generally in the low kHz range. Here, we analyze how the conventional ORT control method limits the topography tracking quality and hence the imaging speed. The closed-loop controller in conventional ORT restricts the sampling rate to the ORT rate and introduces a large closed-loop delay. We present an alternative ORT control method in which the closed-loop controller samples and tracks the vertical force changes during a defined time window of the tip-sample interaction. Through this, we use multiple samples in the proximity of the maximum force for the feedback loop, rather than only one sample at the maximum force instant. This method leads to improved topography tracking at a given ORT rate and therefore enables higher scan rates while refining the mechanical property mapping.

# Introduction

Constant force mode, a widely used AFM imaging mode, utilizes a feedback controller that reads the deflection of the cantilever and keeps the applied tip-sample vertical force at a fixed setpoint value by adjusting the voltage applied to a Z axis nano-positioner. While this AFM mode achieves high precision in controlling vertical forces, the high lateral forces applied while scanning limits the application of this AFM method when gentle and non-damaging imaging is required, for instance on soft biological materials [1]. In order to make the instrument technique suitable for imaging fragile samples, several dynamic modes that rely on the resonance characteristics of the cantilever have been introduced [2], [3]. Although these methods are gentler than contact mode, interpreting and controlling the vertical force exerted on the sample is not straightforward.

To achieve a better tip-sample force control, Rosa-Zeiser et. *al*. [4] presented an off-resonance dynamic mode called pulsed force mode, where force versus distance curves are acquired periodically. The maximum cantilever deflection during one period, corresponding to the maximum exerted force, is sampled and fed into a feedback controller. The tip-sample contact duration is limited and easily tunable compared to the constant force mode, resulting in a significant drop of the applied lateral force [5]. In addition, the acquired force versus distance curves enable extracting sample's physical properties in real time [6]–[8]. For these reasons, derivations of pulse force modes have grabbed the interest of AFM users whose research areas are focused on soft biological samples [9]–[14] or material properties[15]–[20]. AFM companies included variations of the pulse force mode in their microscopes such as PeakForce™ Tapping (Bruker), Digital Pulsed Force Mode™ (WITec), HybriD mode (NT-DMT), and WaveMode (Nanosurf). While these implementations have subtle differences, we refer to these modes (that tap on the surface at a frequency far from cantilevers' resonance frequency) from now on in this manuscript generically as *off-resonance tapping* (ORT) modes.

Despite its many benefits, ORT has the drawback of limited scan speed which stems from the frequency of actuation and the mechanical bandwidth of the scanner and cantilever. In most ORT applications, the piezo that tracks the topography changes is also used to generate a periodic Z-axis motion. However, the resonance frequency of the piezo sets a limit on the actuation frequency. To overcome this problem, one approach is to add a second piezo on Z-axis with a higher resonance frequency [21]. Another approach is direct actuation of the cantilever, as it typically possess higher resonance frequency [22]–[24]. In particular, photothermal actuation, which utilizes laser-induced heating of the cantilever, has led to a significant increase in the achievable ORT frequency [22]. The other speed-limiting factor is the snap-off ringing of the cantilever, especially for applications in air and vacuum [25]. Although this physical phenomenon can be used to extract material properties [26], it slows down the imaging speed since enough time must elapse between snap-off and subsequent contact with the sample so that the cantilever ringing diminishes to avoid sample and/or cantilever damage.

Many attempts have been made on the design of actuators and sensors to speed up ORT techniques. However, the utilized controllers still have the same core scheme as in the early works [27], [28] where only the maximum force is sampled and fed into the feedback controller, and the full potential of ORT is not exploited. Echols-Jones et, al. [29] have shown that increasing the imaging speed in ORT modes is possible by implementing a control algorithm that takes advantage of the tip-sample interaction between the first contact point and the maximum force instant.

Here, we present a detailed analysis of how the sampling rate and delay in the conventional control method in ORT modes intrinsically limit the closed-loop control of the cantilever deflection, and therefore the imaging speed. To overcome these limitations, we introduce an alternative control technique that takes advantage of the non-zero contact duration, enabling an enhanced force control. This method provides rapid control of the maximum force, resulting in better image quality at faster scan rates. Importantly, this method not only preserves the mechanical properties' maps but also provides improved material contrast at higher imaging speeds.

## Results

It is possible to analyze the conventional ORT control framework [27], [30] by decomposing it into its different functional blocks. In the beginning of an ORT cycle, the sample and the tip are separated by an offset distance, and the controller electronics generate a displacement on the Z-axis to establish tip-sample contact (Figure 1 (A)). The displacement is often obtained by actuating the Z piezo with a sinusoidal waveform to avoid exciting the actuator's resonance. After this sinusoidal displacement covers the initial vertical offset distance between the surface and the tip, $Z_{\text{offset}}$, the cantilever starts to deflect (Figure 1 (B)). As most AFMs use controllers implemented in the digital domain, the cantilever deflection is acquired by an analog to digital converter (ADC) to reconstruct periodic force-distance curves (Figure 1 (C)). When the $Z_{\text{offset}}$ changes due to variations in sample´s topography, both the duration of the contact and the maximum force value change. The difference between the measured maximum force and the user defined reference maximum force (setpoint) is the process variable that the feedback controller uses to calculate the error (Figure 1 (D)). Figure 1 (E) shows that a step-up in sample topography (Figure 1 (E) i.), gives rise to an increase in the maximum force value (Figure 1 (E) ii.). The feedback controller within the traditional control scheme only samples the maximum force error by synchronizing with the maximum point of the sinusoidal Z-axis actuation. As illustrated in Figure 1 (E), this operation is equivalent to modulating the continuous time deflection signal Figure 1 (E) ii with an impulse train Figure 1 (E) iii., which is defined in Equation 1, where $T_{ort}$ stands for ORT period.

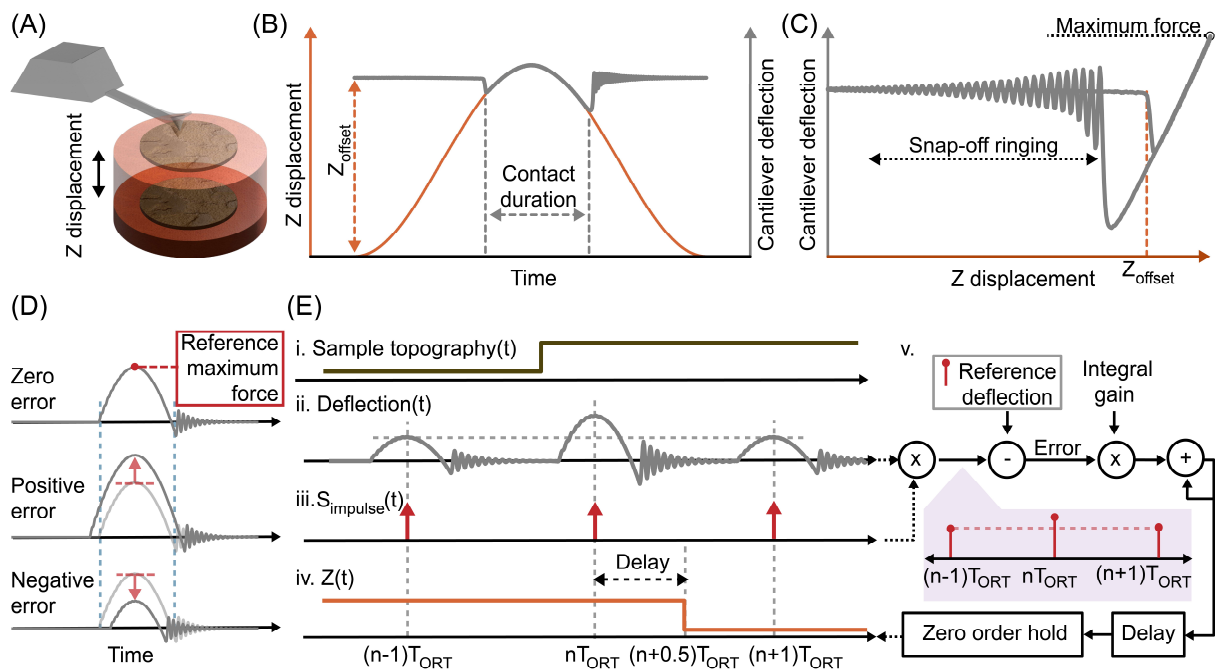

**Figure 1:** (A) In ORT modes, consecutive force versus distance curves are generated with a periodic Z displacement. (B) To avoid exciting the actuator's resonance, a sinusoidal actuation signal is applied, and the cantilever deflection is recorded as a function of time. (C) Cantilever deflection and Z displacement are used to reconstruct the force-displacement curve, where the maximum force can be easily extracted. (D) The difference between the recorded maximum force and setpoint maximum force defines the error for the closed-loop control system. (E) A change in the sample topography (i.) leads to an error in the deflection signal (ii.). The feedback controller acquires the maximum deflection which is subtracted from the reference value to calculate the error. This process can be represented by modulating the deflection signal with an impulse train function (iii.). The Z signal (iv.) that minimizes the error induced by the change in topography is generated by the feedback controller (v.) The feedback controller applies the integral gain to the error and accumulates it with the previous values. The new Z value is applied at the minimum Z displacement instant which is half a period delay after the maximum force instant. The zero-order hold represents the conversion of discrete control iterations into a continuous time signal.

$$S_{impulse}(t) = \sum_{n=-\infty}^{\infty} \delta(t - nT_{ort}) \quad (1)$$

After, the feedback controller calculates the Z output corresponding to the sampled error, and the updated Z value is applied with a delay to ensure that the tip and the sample are no longer in contact. Often, the new Z value is delayed until the instant of the furthest tip-sample distance, which is half a period after the maximum force sampling (Figure 1 (E) iv.). Only at that point, the new Z value is transferred to an analog voltage with a digital to analog converter (DAC). Mathematically, digital to analog conversion can be modeled as a zero-order hold block. Thus, the control process can be represented with a sampler, an integrator, a delay element, and a zero-order hold block (Figure 1 (E) v.).

To analyze the reference tracking quality of the controller, we model a unity gain closed-loop system (further information in the Supplementary Information, Figure S1 A). In order to investigate the frequency response of the feedback controller, we define the closed-loop transfer function between error, E(s), and disturbance, R(s), in the Laplace domain as in Equation 2, where $T_{ort}$, $T_{delay}$, $K_i$ denote the ORT period, the delay, and the integral gain, respectively (derivation of the formula is provided in the Supplementary Information). The non-zero delay in the closed-loop system forms a second-order transfer function which can exhibit a high-amplitude resonance peak in the closed-loop, potentially leading to instabilities, depending on the $T_{ort}$, $T_{delay}$, and $K_i$. Through simulations based on Equation (2) we have determined that the ORT period and the delay impose a limit for the integral gain value (experimental results in Supplementary Information 1, Figure S1 B). In order to prevent closed-loop oscillations caused by the resonance peak, it is necessary to decrease the integral gain, which in turn reduces disturbance rejection and limits the achievable imaging speed.

$$\frac{E(s)}{D(s)} = \frac{sT_{ort} + s^2 \frac{T_{ort} + T_{delay}}{2}}{K_i + s\left(T_{ort} - K_i \frac{T_{delay}}{2}\right) + s^2 \frac{T_{ort} T_{delay}}{2}} \qquad (2)$$

In constant force mode, the feedback controller gets updated at the ADC sampling rate. Typically, the ADC sampling rate of the controller is much higher than that of the ORT rate, and the Z piezo response determines the closed-loop delay. Therefore, the maximum achievable integral gain and disturbance rejection are often limited by the mechanical properties of the Z actuator [31]. In ORT mode, however, the tip and the sample are in contact for a finite time window only. The feedback controller could run for all sampled points within this contact window, similarly to constant force mode (Figure 2 (A)). This method enables the feedback controller to collect multiple sequential samples at the ADC sampling rate, within the limited time window, per ORT interaction. This provides faster feedback iterations compared to the conventional ORT controller that only samples once per interaction period.

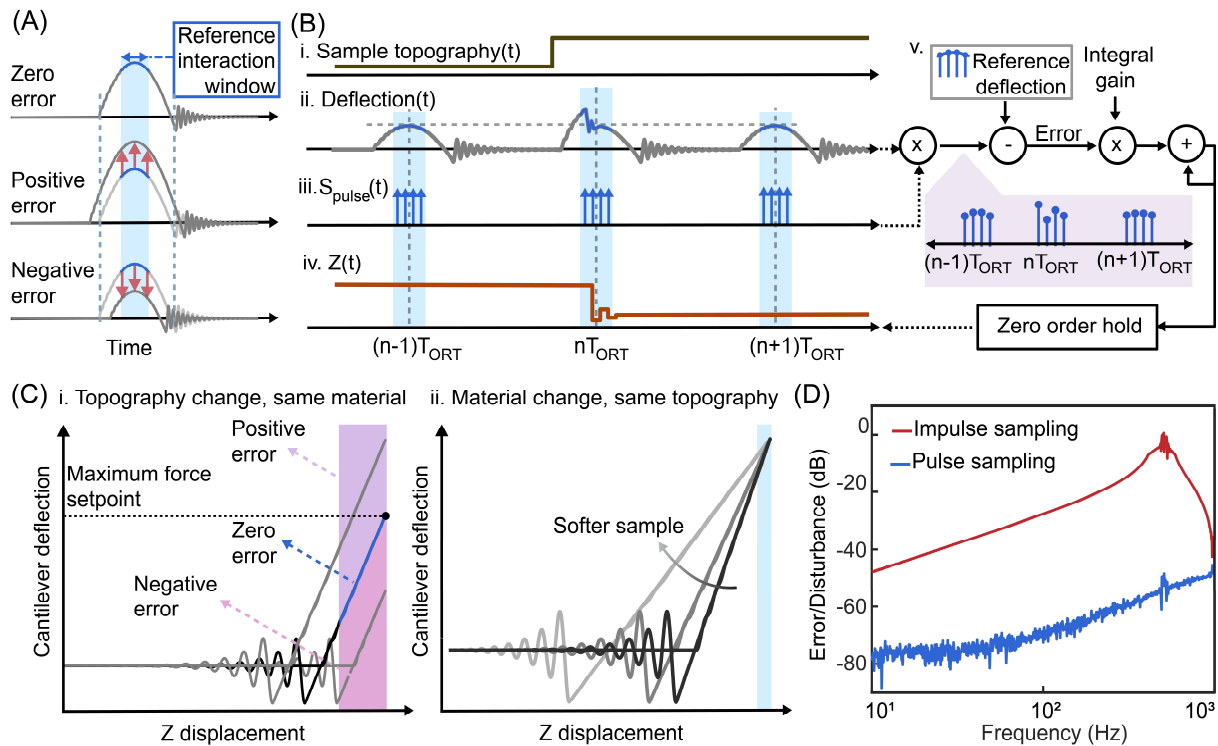

**Figure 2:** (A) Deflection values during the interaction window (shadowed in blue colour), which is selected around the maximum deflection point, are used for error calculation. (B) illustrates the response of the proposed method for a change in topography: i. topography change, ii. cantilever deflection, iii. pulse train sampling function, iv. corresponding Z signal. The Z output of the feedback controller is applied immediately during the selected interaction window to minimize the error within the same ORT cycle. In order to define the reference deflection a previously saved force curve is used. (C) For a sample composed of a single material, a wide interaction curve can be used (i.). If the sample consists of materials with different stiffnesses, a smaller reference window should be selected (ii). Since the Z displacement is sinusoidal, its velocity slows down around the maximum deflection and provides enough time window to collect multiple data points. (D) The proposed method is compared with the conventional method in a unity feedback closed-loop. With 40 dB less error, the proposed method enables more accurate tracking and faster imaging.

Figure 2 (B) illustrates the operation principle of the proposed method for a step-up in the sample topography (i.). The feedback controller samples all the deflection values within the reference interaction window (ii.), which can be modelled as modulating the deflection signal with a pulse train function (iii.). The mathematical expression of the pulse train function is given in Equation 3, where $T_{sample}$ and $T_w$ denote the sampling period of the controller, and the width of the selected reference interaction window, respectively. After a change in topography, the first deflection value of the following interaction window is used to calculate the error, revealing the topography before reaching the maximum force point. The feedback controller generates a Z signal to compensate the error while the tip and the sample are still in contact (iv.). The new Z value (iv.) changes the deflection (iii.), providing the controller an opportunity to measure the error and update the Z position once again. Therefore, by having multiple closed-loop control iterations within the selected $T_w$ window per ORT cycle, this method provides faster and more accurate topography tracking.

$$S_{pulse}(t) = \sum_{m=-\frac{T_w}{2}}^{\frac{T_w}{2}} S_{impulse}(t - mT_{sample}) \qquad (3)$$

Deciding the $T_w$ width of the reference interaction window is a critical step while using the proposed method. When imaging a sample made of a single material, an interaction window covering the full positive force gradient values can be selected (Figure 2 (C) i.). However, in the case of composite samples consisting of different materials, using a reference curve taken on one material would create problems when imaging on a second material, since different materials exhibit different force-curves for the same maximum force setpoint value, as shown in Figure 2 (C) ii. The slope of the force-distance curve depends on the effective spring constant between the tip and the sample. To mitigate this issue, we select a small-time window around the maximum deflection point, where the tip-sample relative velocity is the smallest due to the sinusoidal shape of the periodic Z movement. This approach minimizes the error margin caused by material differences.

The open-source AFM controller developed in our laboratory has provided us with low-level access to implement this proposed method. In real-time, the data points within the selected window are subtracted from a recorded reference interaction curve to calculate the error. The use of a point-per-clock sampling approach for generating the actuation and data acquisition allows for easy indexing and synchronization of the data points.

To compare our control method with the traditional ORT one, we measured the frequency response of error over disturbance in a unity feedback closed-loop (See Supplementary Information 2 for the illustration of the experiment setup.). Experimental results shown in Figure 2 (D) demonstrate that the pulse sampling method rejects the disturbance significantly better, providing 40 dB less error, compared to the impulse sampling method. This indicates that integral gain can be selected at higher values in the pulse sampling method without causing an oscillation as it has less delay and an effective faster sampling rate.

We have tested both methods on a Bruker Multimode AFM head, J-type tube scanner, to verify the improvements of the proposed method. Figure 3 (A) is the schematic of the setup where we placed an additional piezo actuator on top of the tube scanner (for characterization purposes only). The Z piezo of the J-scanner was used to simulate a surface topography with a known disturbance signal, while the feedback controller drove the second piezo to compensate for the disturbance and apply the ORT actuation signal as shown in Figure 3 (B). We selected a 30 Hz triangular waveform (Figure 3 (C)) as the disturbance. Such a triangular disturbance is common in AFM measurements, since it is difficult to avoid the sample tilt in real AFM applications. While a single integral controller ensures zero steady-state error for a step disturbance, for a triangular disturbance, the error is a constant non-zero value which can be used to compare the closed-loop tracking quality. The maximum achievable integral gains before the system becomes unstable were applied in both methods. The ORT period was 1 ms for both methods and the reference window width for the pulse sampling mode was 64 µs, centered around the maximum force (this pulse width was used in all the subsequent experiments presented). The Z signals generated by the controller and the deflection of the cantilever are displayed in Figure 3 (D) for the impulse sampling method, and in Figure 3 (E) for the pulse sampling method. Both methods can track the topography disturbance (Figure 3 (D) i. and Figure 3 (E) i.), however they have a residual steady state error on the increasing and decreasing slope of the disturbance (see the envelope of the maximum deflection points, highlighted purple curves in Figure 3 (D) ii. and (E) ii.). Nevertheless, there is a clear distinction in the amount of this steady state error for the two methods. A closer look at the Z signals in Figure 3 (D) iii. and (E) iii. shows that the closed-loop controller in pulse sampling mode adjusts the Z signal more than once, whereas impulse sampling mode updates the Z only once per ORT cycle. Moreover, the difference between maximum deflection values in Figure 3 (D) iv. and (E) iv. and the setpoint (denoted by the dashed black line) shows that pulse sampling mode can control the maximum force exerted on the surface more accurately. To express the closed-loop error of a ramp disturbance in Laplace domain, we can insert the Laplace transform of a ramp function, (slope)/$s^2$, into Equation (2) as $R(s)$. The steady-state error, $\lim_{s \to 0} sE(s)$, is proportional to the ORT period and the slope, and inversely proportional to the integral gain. Thus, a negative slope indicates a negative error while a positive slope indicates a positive error for an integral gain operation. The deflection signals in Figure 3 (D) ii. and (E) iii. show that the maximum force is always below the setpoint for a negative slope, and exceeds the setpoint for a positive slope. As the steady-state error is inversely proportional to the integral gain, observing less steady-state error in pulse sampling mode ORT is a clear indication of higher integral gain. Consequently, pulse sampling ORT mode provides higher closed-loop bandwidth and faster scan rates by enabling higher integral gain without causing closed-loop oscillation.

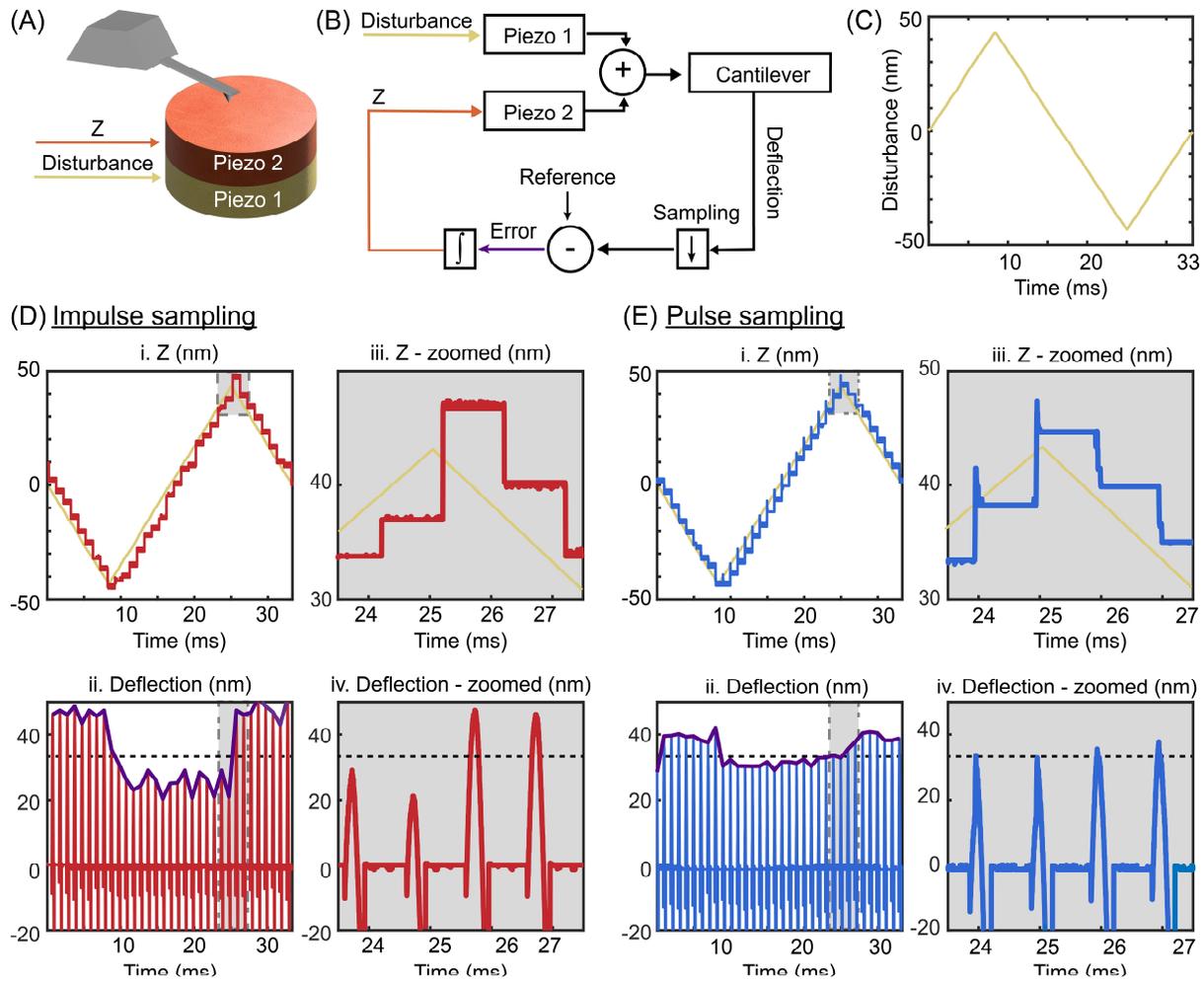

**Figure 3:** The tracking performance of both methods was evaluated in an experiment setup (A), where a scanner equipped with an additional piezo on the Z axis was used to introduce an external disturbance (B), a 30 Hz triangular waveform (C). In panels (D) and (E), the generated Z signal (i.) and deflection of the cantilever (ii.) in one period of disturbance are displayed for impulse sampling and pulse sampling, as well as zoomed windows at the turnaround are displayed (iii., and iv.). The Z response from both methods shows that the closed-loop control can track the external disturbance with discrete iterations (the ideal Z response is denoted with a yellow line). While impulse sampling has only one discrete iteration per ORT cycle, pulse sampling has multiple iterations, helping to minimize the error within a single ORT cycle. The maximum deflection values of each ORT cycle (purple curve) clearly indicate that pulse sampling outperforms impulse sampling in tracking the maximum deflection setpoint (dashed black line).

Both control methods have been tested on a custom-built tip scanner system, comparing their imaging performance. A grid sample (squared pits of 10 µm XY pitch and 200 nm depth) was chosen since it has sharp topography changes on the pit edges and a small tilting angle on the flat surface, providing both high-frequency and low-frequency disturbances to examine the system response. A peak-to-peak ORT actuation of 50 nm at 2 kHz rate was applied to the Z piezo. We have mounted a Fastscan-A (Bruker) cantilever, setting the scan rate to 4 Hz. The acquired image has 250x250 pixels so both trace and retrace images have 1 ORT cycle per pixel. Integral gains in both measurements were set just below the values that would lead to observable oscillations. Height images in Figure 4 (A) show that the pulse sampling method has better tracking performance for rapid topography changes of the pits. The method provides sharper responses at the pit edges, indicating improved performance of the new method for high-frequency disturbances. The line profile in Figure 4 (B) demonstrates that, for the traditional ORT method, the tip loses contact with the surface and re-establishes contact only after a few ORT cycles (known as parachuting). In pulse sampling mode, the duration of parachuting is reduced compared to traditional ORT. The error (ii) on the rising edges shows a significant drop of the exerted force for the pulse trace compared to the impulse trace. Moreover, a substantial decrease in the constant error on the flat region due to the tilting angle of the sample is evident from the line profiles in Figure 4 (C). The RMS values of the error line profiles in Figure 4 (C) ii. (trace + retrace), are 29 nN$_{rms}$ and 3.9 nN$_{rms}$ for impulse and pulse sampling methods, respectively. Compared to the setpoint (67 nN), the impulse sampling method leads to a constant force error of around 40% of the nominal setpoint, whereas this value was lower than 6% for the pulse sampling method. This indicates that also at lower disturbance frequencies, the pulse sampling method minimizes the error more effectively. The proposed ORT control method is more effective in not only tracking high frequency disturbances, but also improving tracking of lower frequency components of the sample topography.

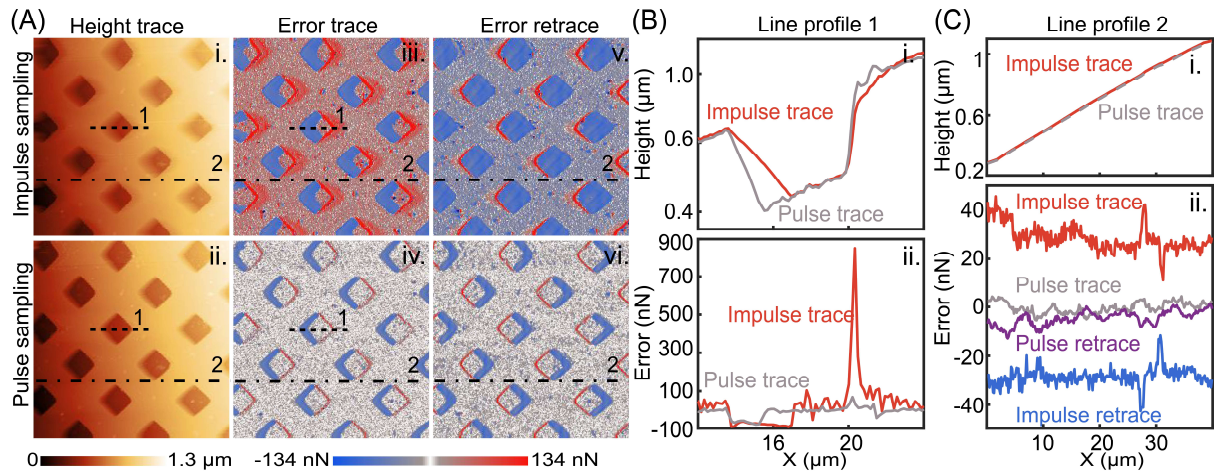

**Figure 4:** A grid sample was scanned with both methods to compare the image quality and the tracking performance. (A) Pulse sampling not only improves tracking for high frequency disturbances such as pit edges but also for low frequency disturbances such as a flat surface that has a certain slope. (B) Line profile 1 shows that pulse sampling reacts faster to rapid topography changes of the pit edges, resulting in significantly less error. On the falling edges the tip loses the contact and the error is equivalent to the negative setpoint. In the pulse sampling mode the tip-sample contact is re-established faster. On the rising edges, the cantilever reaches the contact earlier and deflects more. The pulse sampling mode provides around 10 times less error as it reacts before reaching the maximum deflection point. (C) Line profile 2 demonstrates that in pulse sampling there is a singficant decrease in steady state error of the sample tilt.

Extracting mechanical properties from the interaction curves is one of the most powerful capabilities of ORT techniques. Together with the topographical image, they provide material contrast mapping of the sample in real time. However, the topography feedback error should be minimized to reduce the effect of sample topography on the mechanical property measurements. For example, dissipated energy on the sample, simply the integral of the area between the approach and the retract curves, depends on the applied maximum force. Therefore, the maximum force error affects the measured dissipation values. To get better material contrast, one usually scans slowly to minimize the error. A lower imaging speed decreases the frequency of the disturbances induced by the topography, for which the controller has better tracking performance. It also reduces the error per pixel as there are more data points to average. Achieving less error for a given speed and topography, the pulse sampling method better minimizes topography artefacts on material contrast images. To demonstrate this, we imaged a blend of polystyrene and low-density polyethylene (PS/LDPE) sample (SPM LABS LLC., Tempe, USA), mounting a RTESPA-300 cantilever (Bruker), and using both methods with the same ORT parameters, 50 nm peak to peak amplitude and 2 kHz ORT frequency. Figure 5 shows 250x250 pixels images captured at different scan rates with both methods. Panel (A) displays the height image captured at 0.5 Hz line rate with the pulse sampling method. Panel (B) displays the error channels at 0.5 Hz line rate (i., ii.) and 1 Hz line rate (iii., iv.), for impulse sampling mode (i., iii) and pulse sampling mode (ii., iv.). Similarly, panel (C) shows the dissipation channels at 0.5 Hz line rate (i., ii.) and 1 Hz line rate (iii., iv.), for impulse sampling mode (i., iii.) and pulse sampling mode (ii., iv.). In the error images, pixels with values outside of the range of +/-25 % of the setpoint (270 nN) are highlighted in red. The projected surface area ratio of the red-colored area to the whole image is calculated to demonstrate the improvement of the pulse sampling method. For the impulse sampling mode, the calculated ratios are 8.1% and 32% for 0.5 Hz and 1Hz line rates, respectively. For the pulse sampling mode, calculated ratios are 0.2% and 2.7%, showing the significant improvement in tracking the topography. Moreover, the highlighted pixels in the error image also produce artefacts on the dissipation images (C). As predicted, more accurate topography tracking of the pulse sampling mode yields less error, therefore, the dissipation channel shows fewer topography related artefacts than the impulse sampling.

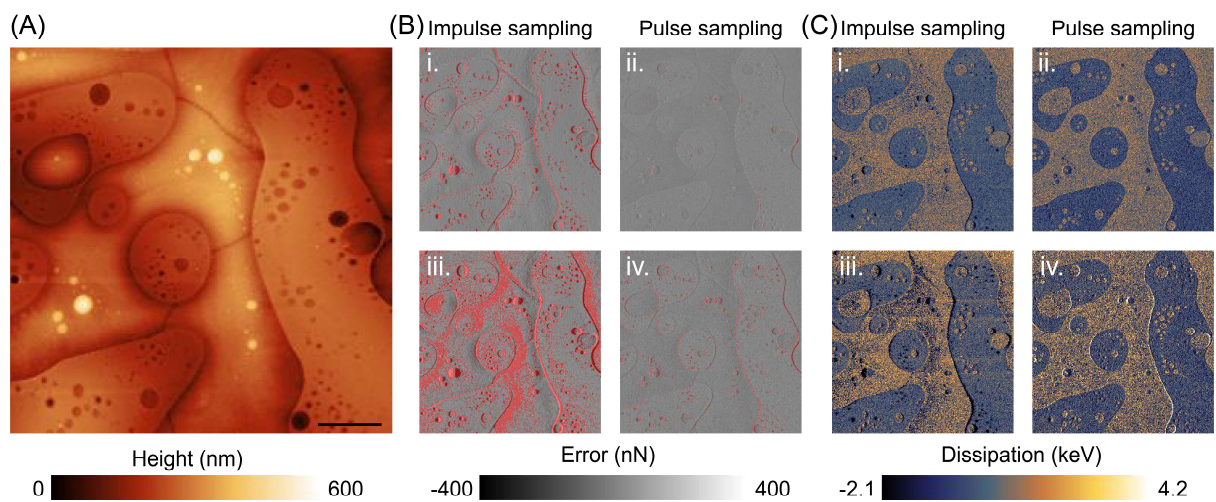

**Figure 5**: Both methods are used to image PS/LDPE. The height image of the sample in panel (A) was captured using the pulse sampling mode imaging at 0.5 Hz line rate. Panel (B) displays the error channels obtained with both methods at the line rates of 0.5 Hz (i., ii) and 1 Hz (iii., iv.). Panel (C) presents the corresponding dissipation channels. To better highlight the differences on the error channels and their impact on the dissipation channels, pixels with values more than 68 nN and less than -68 nN are colored as red in (B). The error images show that the pulse sampling method tracks the topography better, as it results in lower error values. The poor tracking of the impulse sampling method leads to the appearance of topographical artefacts in the dissipation channel, which becomes more pronounced as the imaging speed increases. In contrast, better tracking performance of the pulse sampling method improves the material contrast at the same surface speeds. Scale bar 10 μm.

## Discussion

The improvement of the proposed method over the conventional method relies on the selected width of the reference interaction window and the response time of the Z piezo. The width of the interaction window determines the number of closed-loop control iterations within a single ORT cycle. A wider interaction window allows for more iterations, potentially improving the tracking quality. The width of the interaction window is influenced by the ORT actuation amplitude and the setpoint. A smaller ORT actuation amplitude and a higher setpoint enable the use of a wider reference interaction window by prolonging the tip-sample contact duration. However, widening the interaction duration eventually leads to the system converging toward contact mode, resulting in increased lateral forces. Therefore, a trade-off between the scan speed and lateral force needs to be considered. The bandwidth of the Z actuator is the other factor that limits the achievable improvement with the pulse sampling method. To fully benefit from multiple closed-loop iterations per ORT cycle, the Z piezo should be sufficiently fast to respond within the same interaction window. It is ineffective to have a large number of data points per ORT cycle if the Z piezo is too slow, since the controller would calculate the same error for all data points as the piezo is too slow to respond to the required fast changes. When the ORT rate is high, close to the resonance frequency of the Z piezo, or when the direct actuation of the cantilever is used to get high ORT rates, the Z piezo becomes too slow to adjust the multiple corrections per ORT cycle. On the other hand, the pulse sampling mode provides better tracking compared to the conventional method, as it does not have the intrinsic delay. We observed the marginal improvement in tracking when we applied the method to high-speed photothermal ORT. Moreover, since the proposed method provides better tracking at lower ORT frequencies, it becomes possible to achieve faster scan rates at lower ORT rates. This is particularly advantageous in air or vacuum where slower ORT rates need to be chosen to circumvent the effect of snap-off ringing.

Another important aspect to consider in the implementation of the proposed method is the challenge posed by non-equally spaced consecutive error points. This issue arises due to the presence of two different time constants, requiring the use of two different integral gains in our discrete-time accumulator implementation. The data points within the reference window are equally spaced by the sampling period of the ADC. However, this does not hold outside the reference window in which there is a larger time interval between the last data point of one reference window and the first point of the next one. This interval-time difference depends on the ORT period and the width of the reference window. This means that different gain values can be applied on the system without violating the stability. Consequently, the first error value of the interaction window should be multiplied by a higher integral gain compared to the others. We implemented two different independent gains in our controller for the above presented experiments. Although this introduces an additional integral gain parameter to adjust, it allows for individual tuning of gain values, providing more flexibility. It would be possible to use only one integral gain, linking the second one by the ratio of the two time constants.

In the current implementation of the proposed method, the reference curve is calculated by averaging multiple recorded waveforms to reduce the noise level. However, this reference only represents the tip-sample interaction for one given material, which can lead to over or under correction of the interaction when the tip touches a different material. In future implementations, multiple reference curves could be recorded for different materials of the sample. As the slope of the force-distance curve during approach reveals the material contrast, the controller can automatically select the correct reference curve for the probing point before reaching the reference interaction window.

## Conclusion

ORT modes offer the advantage of gentle probing and the ability to perform multiparametric imaging. However, the sampling rate of the feedback control significantly decreases to the kilohertz range compared to the megahertz range in contact mode. In this study, we theoretically and experimentally demonstrated how the conventional ORT control method is limited in terms of closed-loop tracking and imaging speed due to the down-sampling and the introduced hold delay. We propose a new ORT control method that takes advantage of non-zero deflection values in the contact window and run the closed-loop control at contact mode feedback rate for a short width pulse. Theoretical and experimental results demonstrate that the deviation from the reference maximum force due to the topography changes is reduced by this method. This minimizes sample damage and tip wear. As a result, the new method enables faster imaging and better material contrast in mechanical property images.

## Supporting Information

1. Simulations on unity feedback closed-loop ORT controller

Figure S1 A illustrates the controller in a unity gain closed-loop feedback. The integral gain ensures zero steady-state error and, together with the delay element, forms a second-order transfer function which we provide in-detail analysis of in this manuscript.

For simplicity, our theoretical calculations and experimental verification include only the integral gain but not the proportional gain, even though propotional gain would be helpful for better tracking. We have performed theoretical simulations on MATLAB based on Equation 2 which is derived by the following set of equations.

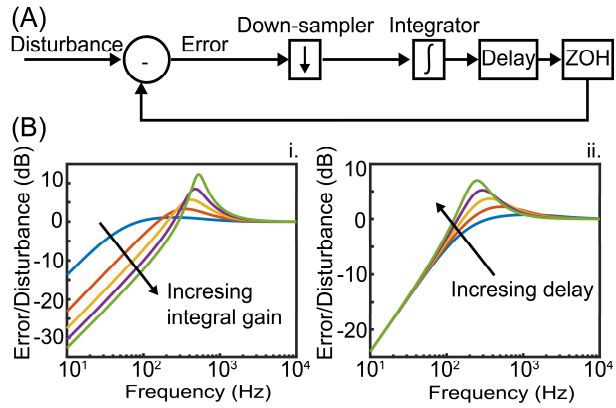

**Figure S1:** The effect of I gain and delay on a simplified ORT controller (A). Error rejection is plotted for 1 kHz ORT rate with constant delay – increasing gain (B) i. and constant gain – increasing delay (B) ii.

discrete domain integrator (accumulation):

$$\frac{k_i}{1-z^{-1}} \tag{s1}$$

discrete time operator:

$$z = e^{sT_{ort}} \tag{s2}$$

delay element:

$$e^{-sT_{delay}} \tag{s3}$$

zero order hold:

$$\frac{1-e^{-sT_{ort}}}{sT_{ort}} \tag{s4}$$

open loop transfer function:

$$G(s) = \frac{k_i}{1-e^{-sT_{ort}}} \times e^{-sT_{delay}} \times \frac{1-e^{-sT_{ort}}}{sT_{ort}} = \frac{k_i}{sT_{ort}} e^{-sT_{delay}} \tag{s5}$$

first order Padé approximation:

$$e^{-sx} = \frac{1-\frac{sx}{2}}{1+\frac{sx}{2}} \tag{s6}$$

Simplification of the open-loop transfer function:

$$G(s) = \frac{k_i}{sT_{ort}} \times \frac{1-\frac{sT_{delay}}{2}}{1+\frac{sT_{delay}}{2}} \tag{s7}$$

closed-loop transfer function:

$$H(s) = \frac{1}{1+G(s)} = \frac{1}{1+\frac{k_i}{sT_{ort}} \times \frac{1-\frac{sT_{delay}}{2}}{1+\frac{sT_{delay}}{2}}} = \frac{sT_{ort}+\frac{s^2}{2}T_{ort}T_{delay}}{k_i + s\left(T_{ort}-\frac{k_iT_{ort}}{2}\right)+s^2\left(\frac{T_{ort}T_{delay}}{2}\right)} \tag{s8}$$

The frequency response of error versus the external disturbance plot displayed in Figure S1 (B) shows the effect of the delay and the integral gain on the resulted error signal. In the simulations depicted on the left panel, we increase the integral gain keeping the ORT period and delay constant at 1 ms and 0.5 ms, respectively. Increasing the integral gain decreases the error for a given frequency. However, it also increases the magnitude and the quality factor of the resonance peak due to the delay. Especially, in slow ORT applications, the zero-order hold delay of half an ORT period is the dominating element in the overall delay of the closed-loop system. On the second simulation set, right panel in Figure S1 (B), we increase the delay by keeping the integral gain constant. Increasing the delay does not significantly change the disturbance rejection at low frequencies, but it creates a resonance peak with a higher amplitude, as seen in Figure S1 (B)ii. To prevent triggering closed-loop oscillations due to a resonance peak, we have to decrease the integral gain and hence degrade disturbance rejection. Simulations based on Equation (1) show that the ORT period and the delay of zero order hold impose a limit for the I gain, which reduces tracking quality and forces the users to image slower.

2. Closed-loop bandwidth measurements of both methods

The functionality of the proposed method is assessed in a test framework, where we have conducted disturbance rejection measurements using a lock-in amplifier (Anfatec eLockIn 205), as illustrated in Figure S2. The reference output of the lock-in amplifier is added to the Z output of the controller as the disturbance and fed to the controller as the deflection input. This is equivalent to replacing the physical AFM with a unity gain component, where the lock-in amplifier reference output represents the topography of the sample. The ORT rate is set at 1 kHz and the interaction window is selected as 12.8 % of the period. Integral gains are set at the highest possible value such that the system does not experience an oscillatory behavior. The error that is recorded during the interaction window is sent to the lock-in amplifier as the input signal. The frequency response of the system is generated by sweeping the reference frequency of the lock-in amplifier and recording the corresponding error, as plotted in Figure 2 (D). For the impulse sampling mode, the remaining error peaks around 500Hz, which is the Nyquist frequency due to the 1kHz ORT rate. Using a higher I gain causes an increase in the amplitude of the peak which triggers an oscillation in the closed-loop system. However, for the pulse sampling mode, the effective integral gain can be set higher due to the significantly lower delay time in the closed-loop, reducing around 40 dB the error for the same disturbance. This experiment demonstrates that employing multiple sampling points per ORT curve and reducing the delay leads to improved tracking performance.

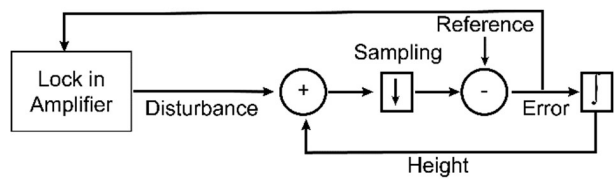

**Figure S2:** Tracking quality of the closed-loop controller is measured by the ratio of recorded error and introduced disturbance with the illustrated setup. Reference output of the lock in amplifier is added to the height signal generated by the controller as the disturbance. The calculated error in the closed is sent to the lock in amplifier as the input.

## Funding
This work was funded by the H2020-UE Framework Program for Research & Innovation (2014-2020); ERC-2017-CoG; InCell; through project number 773091., the Swiss Innovation Agency, Innosuisse, through grant number 36938.1 IP-EE, the Swiss National Science Foundation through grant 200021_182562, and the ETH Domain Open Research DATA EPFL Program